\documentclass{PoS}

\title{Commissioning of CMS Forward Hadron Calorimeters with Upgraded Multi-anode PMTs and  $\mu$TCA Readout}

\ShortTitle{Commissioning of CMS Forward Hadron Calorimeters}

\author{\speaker{Emrah Tiras}$^{1}$, {Burak Bilki}$^{1}$$^{,2}$, {Yasar Onel}$^{1}$\\

\\
$^{1}$University of Iowa, Iowa City, IA, USA\\
$^{2}$Beykent University, Istanbul, Turkey\\

\\
On behalf of the CMS Collaboration\\
\\
University of Iowa, Iowa City, IA, USA\\
E-mail: \email{emrah-tiras@uiowa.edu}}

\abstract{The high flux of charged particles interacting with the CMS Forward Hadron Calorimeter PMT windows introduced a significant background for the trigger and offline data analysis. During Long Shutdown 1, all of the original PMTs were replaced with multi-anode, thin window photomultiplier tubes. At the same time, the back-end electronic readout system was upgraded to $\mu$TCA readout. The experience with commissioning and calibration of the Forward Hadron Calorimeter is described as well as the $\mu$TCA system. The upgrade was successful and provided quality data for Run 2 data-analysis at 13 TeV.}

\FullConference{38th International Conference on High Energy Physics\\
		3-10 August 2016\\
		Chicago, USA}

\begin{document}

\section{Introduction}
The increasing luminosity and resulting radiation conditions entail upgrades of detectors at the Large Hadron Collider (LHC). The upgrade of the Compact Muon Solenoid (CMS) detector includes upgrading many sub-detectors such as Forward Hadron (HF) Calorimeter in the eta region, 3<$\eta$<5 \cite{Collaboration2011}. The HF calorimeter is made of steel absorbers and quartz fibers as active elements to generate photons as particles traverse. These photons are collected with 1728 photomultiplier tubes (PMTs). 

\begin{figure}[h]
\centering 
\includegraphics[width=0.75\textwidth]{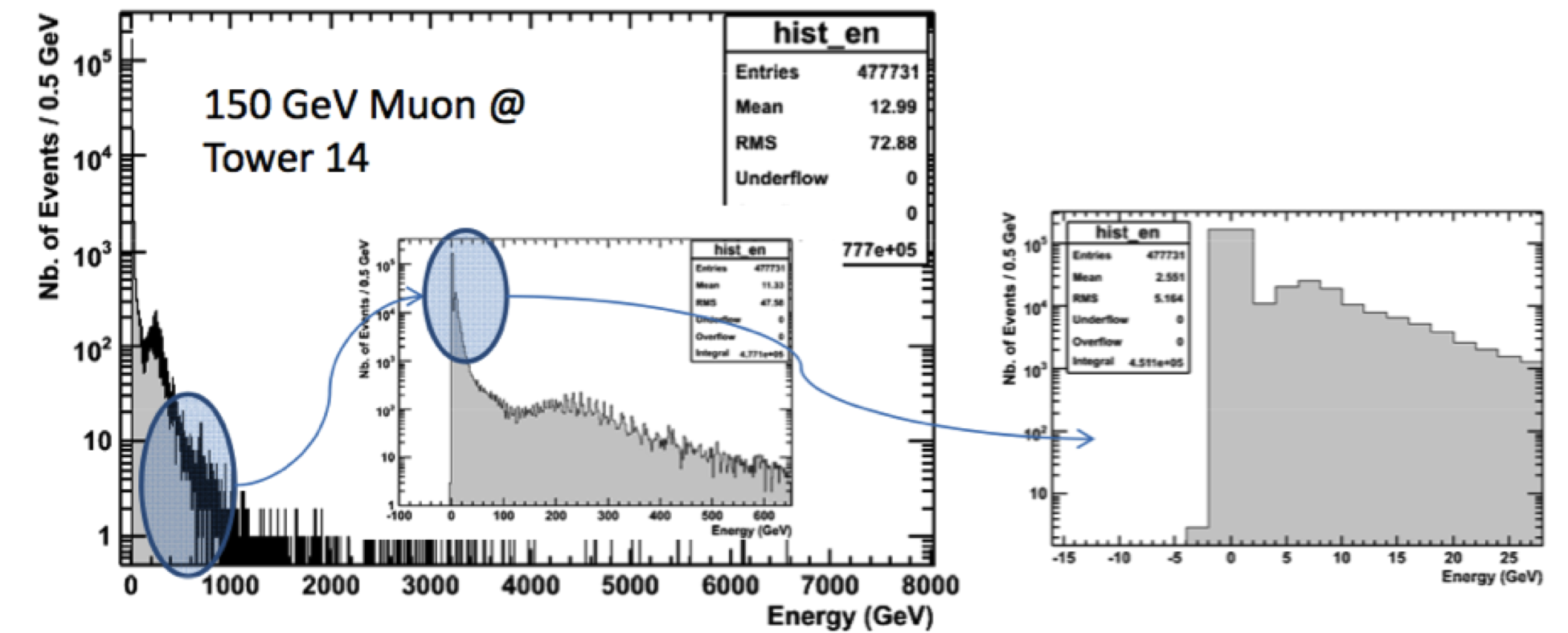}
\caption{Muon response of HF R7525 PMTs at 150 GeV muon test beam.}
\label{LargeEnergyEvents}
\end{figure}

The previous single-anode Hamamatsu R7525 PMTs were decided to be replace with new multi-anode, metal package Hamamatsu R7600U-200-M4 PMTs because of the large energy events detected by the former PMTs \cite{Collaboration2011, HCALCollaboration2010}. The source of the large energy events is high energy muons produced in pp collisions, cosmic particles, and charged particles from late showering hadrons; which produce  $\check{C}$erenkov light at the window of the PMTs. Figure \ref{LargeEnergyEvents} shows the muon response of HF R7525 PMTs at 150 GeV muon test beam during 2004-2008. In the plot, 200 GeV peak is due to muons interacting with the PMT windows and the sharp peak around 7 GeV is due to muons interacting with the HF calorimeter itself. This confirms that the glass window of the previous R7525 PMTs are susceptible to the muons and they needed to be replaced. 

In addition to PMT replacement, the back-end electronic readout system was upgraded to Micro Telecomunications Computing Architecture ($\mu$TCA) readout. Also, 72 readout boxes (RBXs) with new cable design were upgraded and calibration and monitoring systems were improved \cite{Collaboration2012}. 

\section{Characterization of Multi-Anode PMTs}
New multi-anode R7600U-200-M4 PMTs have four anodes, thinner window (< 1 mm) and a metal envelope. Their characterisations at 800 V are high quantum efficiency (38 - 39 \%), high gain (> $10^6$) , low dark current (< 1 nA) and fast timing (rise time < 2 ns, transit time < 25 ns and pulse width < 15 ns) \cite{PMT2014}. They provide many improvements. Thin glass window reduces the signal size of window hit events. Multi anodes allow tagging window hit events and correct the energy of such events. Higher quantum efficiency and gain improve the resolution of the calorimeter and the dynode structure makes PMTs less susceptible to magnetic field. 

\begin{figure}[h]
\centering 
\includegraphics[width=0.43\textwidth]{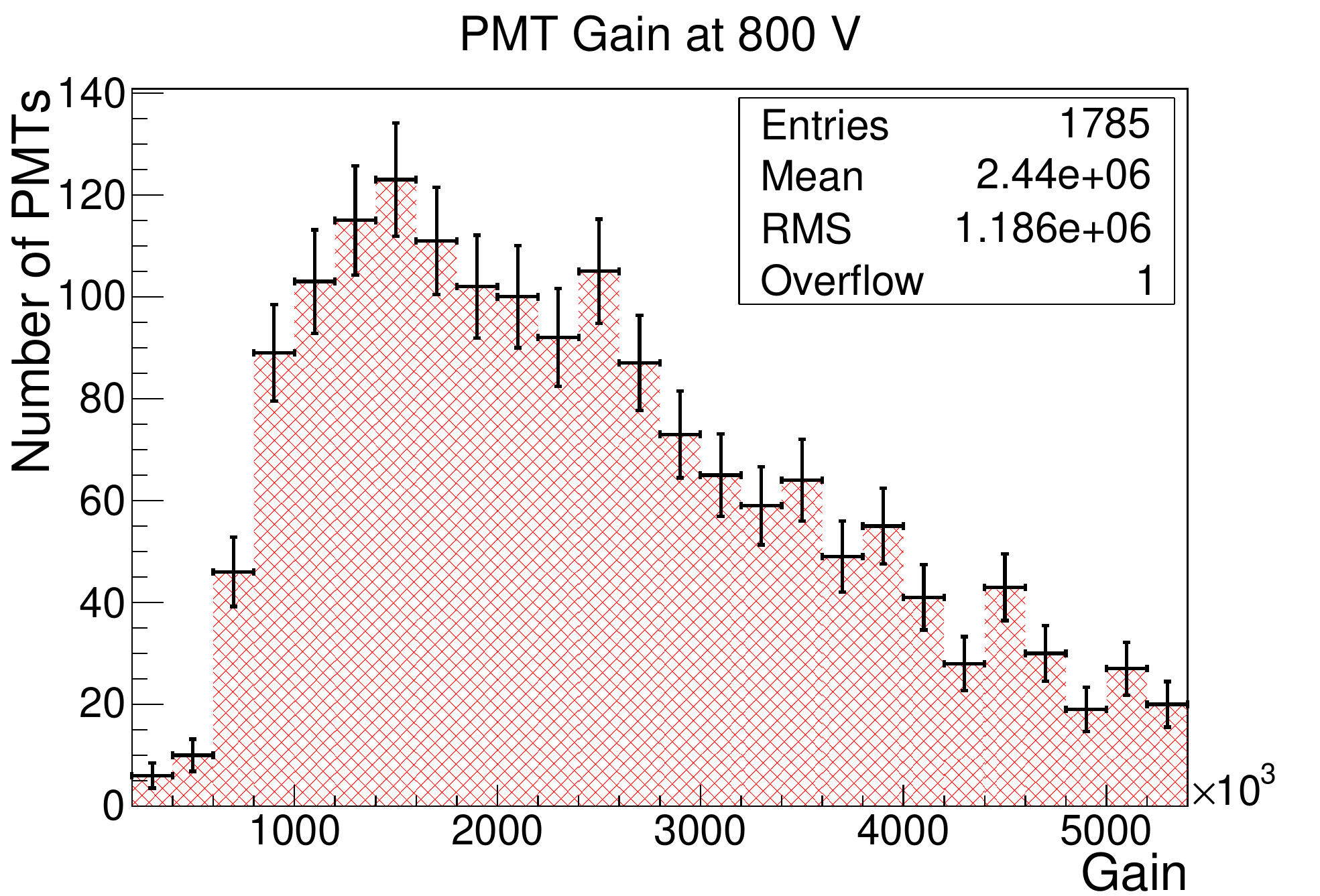}
\includegraphics[width=0.43\textwidth]{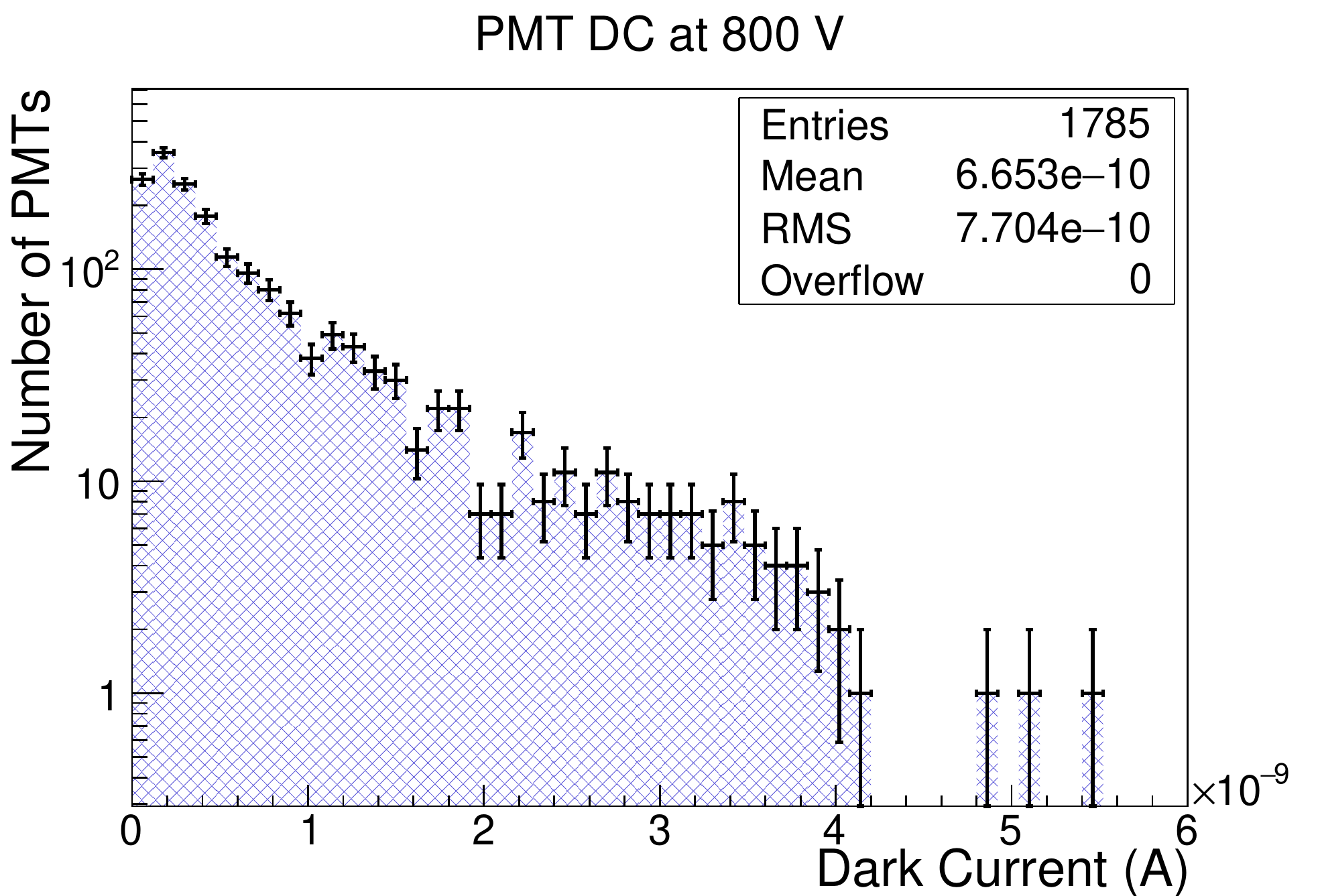}
\caption{Gain (left) and dark current (right) distributions for R7600U-200-M4 PMTs at 800 V. }
\label{GainandDC}
\end{figure}

The characterization tests of new PMTs were done at the University of Iowa PMT test station. All the tests were performed in light-tight boxes. Each PMT was tested for gain, dark current, and timing at the range of operating voltages 600 - 900 V in increments of 50 V.  Figure \ref{GainandDC} shows the gain (left) and dark current (right) distributions of R7600U-200-M4 PMTs at 800 V. The mean (RMS) gain and dark current respectively are $2.44\times10^6$ $(1.19\times10^6)$ and 0.665 (0.770) nA. 

\begin{figure}[h]
\centering 
\includegraphics[width=0.43\textwidth]{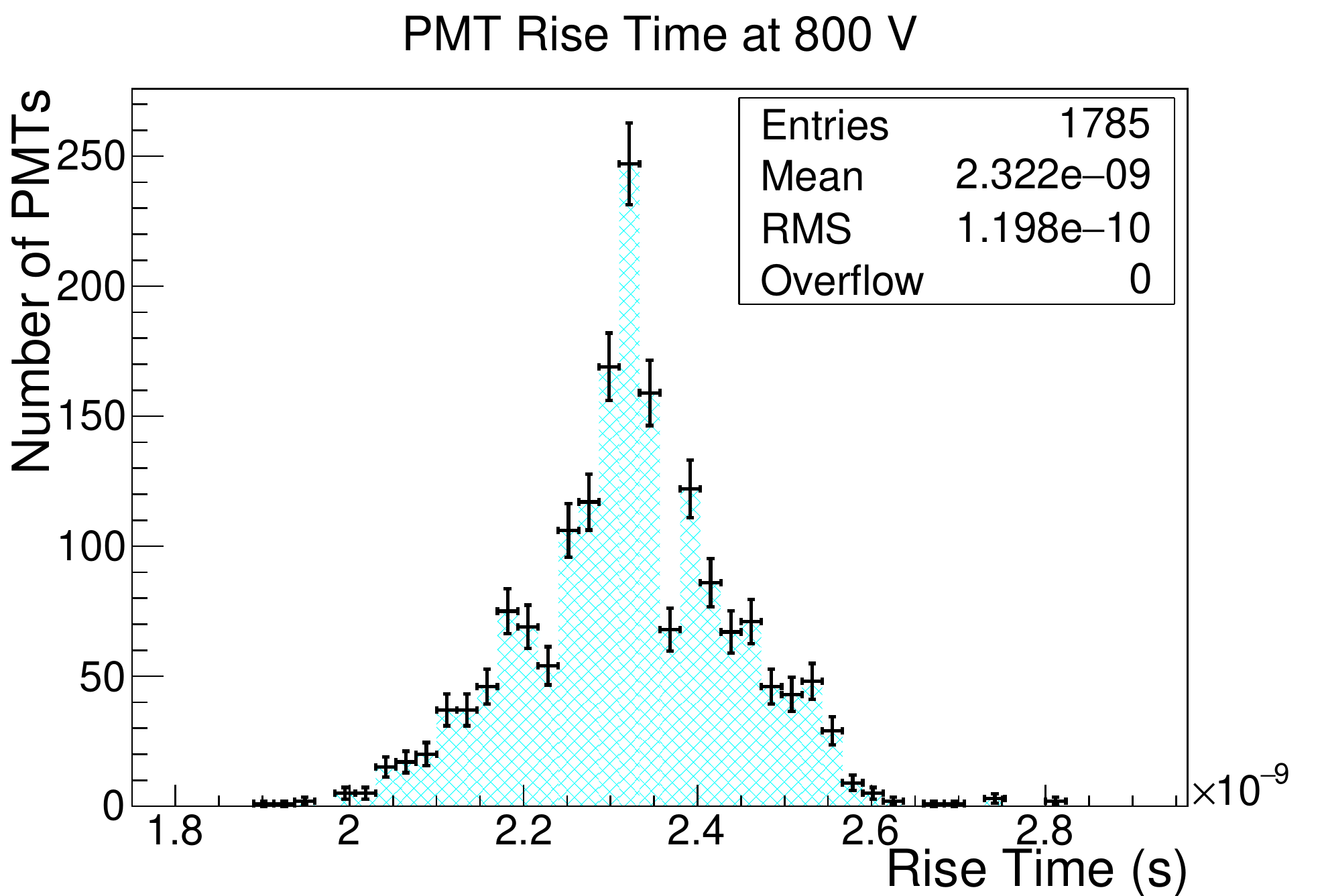}
\includegraphics[width=0.43\textwidth]{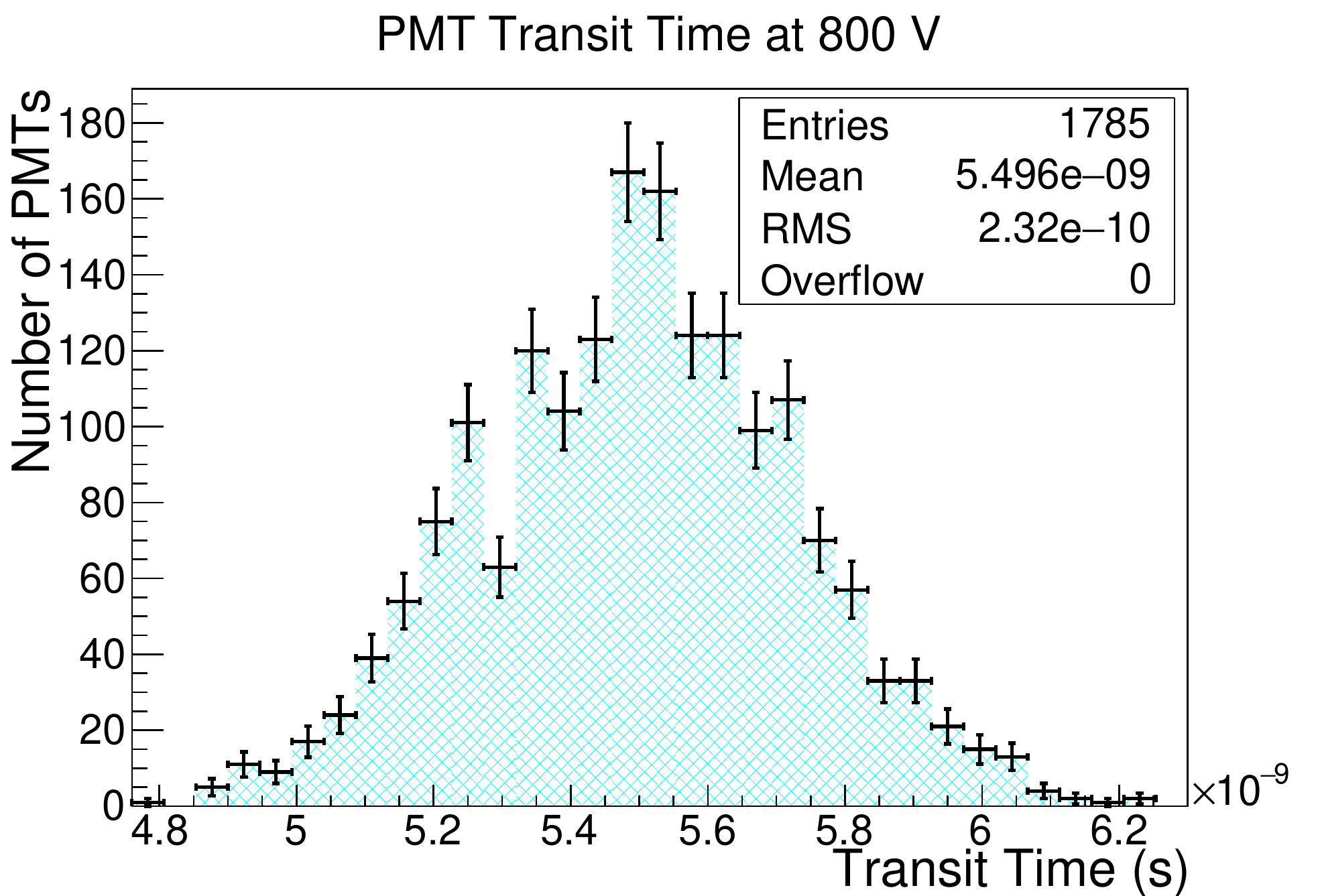}
\includegraphics[width=0.43\textwidth]{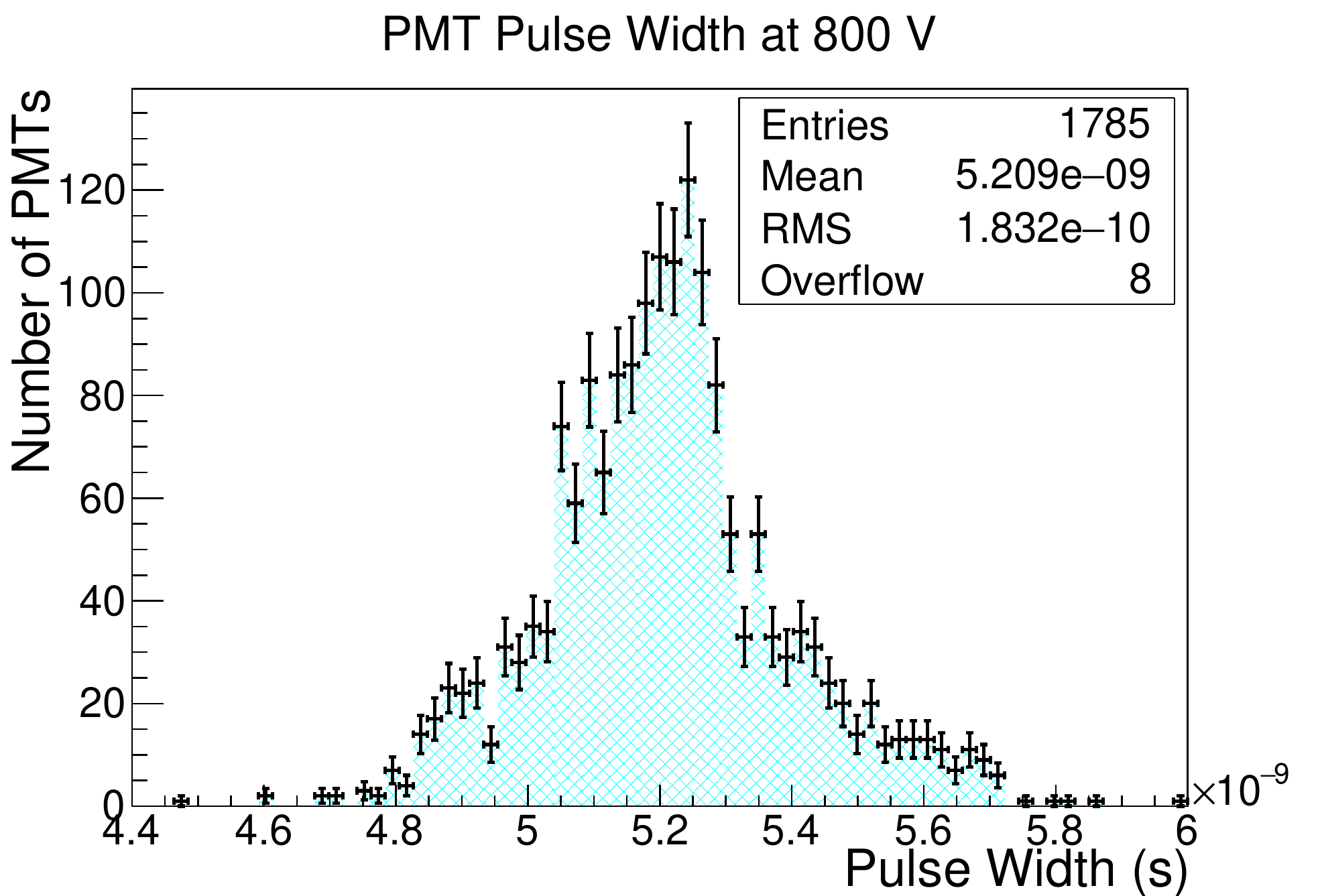}
\caption{Rise time (top left), transit time (top right) and pulse width (bottom) distributions for R7600U-200-M4 PMTs at 800 V.}
\label{Timing}
\end{figure}

Figure \ref{Timing} shows rise time (top left), transit time (top right) and pulse width (bottom) distributions of R7600U-200-M4 PMTs at 800 V.  The mean (RMS) of rise time, transit time and pulse width are respectively, 2.32 (0.12) ns, 5.50 (0.23) ns, 5.21 (0.18) ns. 

\section{Front-end and Back-end Electronic Replacement} 
\begin{figure}[h]
\centering 
\includegraphics[width=0.75\textwidth]{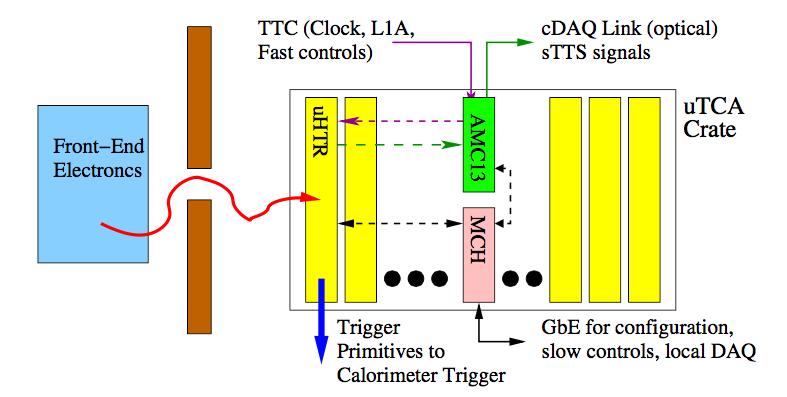}
\caption{Crate layout structure of the $\mu$TCA-based back-end electronics.}
\label{mTCAElectronics}
\end{figure}

New back-end electronics are based on $\mu$TCA system. Figure \ref{mTCAElectronics} shows the crate layout structure of the $\mu$TCA-based back-end electronics \cite{Collaboration2012}. The new modules allow an increased data transfer rate; can receive data via a high-speed 5 Gbps asynchronus link and record histograms with LHC bunch crossing time resolution. 

The front-end electronics will also be replaced during the Extended Year-End Technical Stop of 2016 - 2017. The charge integrator and encoder (QIE) will be upgraded to a new version, QIE10, which provides wider dynamic range, in between 3 fC and 330 pC, and a dead-timeless integration and digitization of charge in 25 ns buckets.

\section{Conclusions}
The CMS Forward Hadron Calorimeters have been going through a photodetector, as well as front-end and back-end electronics upgrade. The vast majority ($\sim$99\%) of new multi-anode PMTs have low dark currents, below 1 nA, and have high gain, above $10^6$. They show fast timing characteristics with a rise time of less than 3 ns, transit time and pulse width less than 6 ns. They operate stably and are now being used to collect data at the CMS experiment. These PMTs with all the improved specifications overcome large enery window event issues in the HF calorimeter, reduce fake background and make the experiment more efficient. New back-end electronics based on $\mu$TCA provide increased data transfer. Also, upgrade front-end electronics, QIE10, will provide wider dynamic range and incorporate TDC functionality.

\end{document}